\documentstyle[preprint,revtex]{aps}
\begin{document}
\draft
\begin{title}
The Kagome Antiferromagnet with Defects: Satisfaction,
Frustration, and Spin Folding in a Random Spin System
\end{title}
\author {E.F.~Shender,$^{(1),}$\cite{ref:a}
V.B~Cherepanov,$^{(2),}$\cite{ref:b} P.C.W.~Holdsworth$^{(3)}$,
and A.J.~Berlinsky$^{(1)}$}
\begin{instit}
$^{(1)}$Institute for Materials Research,
McMaster University, Hamilton, Ontario, Canada, L8S~4M1
\end{instit}
\begin{instit}
$^{(2)}$Institute for Theoretical Physics, University of Manitoba, Winnipeg,
Canada, R3T~2N2,
\end{instit}
\begin{instit}
$^{(3)}$Laboratoire de Physique, Ecole Normale Superieure de Lyon,
France, 69364.
\end{instit}
\begin{abstract}

It is shown that site disorder induces noncoplanar states,
competing with the thermal selection of coplanar states, in the
nearest neighbor, classical kagome Heisenberg antiferromagnet
(AFM). For weak disorder, it is found that the ground state
energy is the sum of energies of separately satisfied triangles
of spins.  This implies that disorder does not induce
conventional spin glass behavior. A transformation is presented,
mapping ground state spin configurations onto a folded triangular
sheet (a new kind of ``spin origami'') which has
conformations similar to those of tethered membranes.

\end{abstract}

It is well known that geometrical frustration in some
non-bipartite lattices prevents long range magnetic order from
being established and allows novel kinds of low temperature
magnetic states to develop\cite{ref:1,ref:2,ref:3}.  The Heisenberg kagome
antiferromagnet with nearest neighbor couplings is one of the
most interesting of such systems. The classical system exhibits a
rich, nontrivial ground state degeneracy, with
both coplanar and noncoplanar states in the degenerate manifold.
For the coplanar states, linear spinwave theory yields one
zero-energy mode for every point in the Brillouin zone\cite{ref:4,ref:5}.
All noncoplanar states have fewer zero modes, and, as a result,
thermal effects select a nematic-like coplanar ground
state\cite{ref:5}, an example of the ``order by disorder''
effect\cite{ref:6,ref:7}. Numerical studies have confirmed the tendency
for thermal selection of the nematic-like state\cite{ref:5,ref:8}, and
there is also evidence\cite{ref:4,ref:5,ref:8,ref:9,ref:sachdev} for a
tendency toward $\sqrt{3}\times\sqrt{3}$ ordering in the plane.

By far the best-studied experimental kagome system is the
magnetoplumbite, $\rm SrCr_{9p}Ga_{12-9p}O_{19}$\cite{ref:10}.
For p=1, this system contains dense kagome layers, separated by
dilute triangular layers, of Cr.    Although its Curie-Weiss
temperature, $\rm \Theta_{CW}$ (for $p=1$) is over 500K, no
sublattice ordering is found down to helium temperature, where a
spin glass, rather than an ordering transition is observed at a
temperature $\rm T_f$.  The ratio $\rm \Theta_{CW}/T_f$ is about
130, at least for $\rm p > 0.5$ \cite{ref:10,ref:ram2,ref:11}. $\rm
T_f$ itself varies rapidly with doping \cite{ref:ram2,ref:11},
having its {\it maximum} value of about 4K near $\rm p=1$,
where one might expect structural disorder to be least important,
and falling monotonically as p is reduced. These observations
raise two questions: (1) Why is spin glass behavior, with a
temperature scale of order J, not generated by non-magnetic
impurities at the 10 to 20\% level, and (2) what is the origin of
the spin-glass-like behavior which is observed even for $\rm
p\approx 1$? It is the first question which is addressed in
this Letter, while the second is discussed briefly in our
conclusions. Our main results are as follows:

\noindent 1. Quite generally we find that disorder induces
noncoplanarity in the ground state. At low temperatures, the
nematic-like state, which is selected by thermal fluctuations, is
overwhelmed by this tendency of disorder to induce noncoplanarity.
\cite{ref:henley}

\noindent 2. For a large class of distributions of spins of random
magnitude, including dilute distributions of
vacancies, the ground state configuraton is such that the energy
of each separate triangle is minimized. We call this the ``Rule
of Satisfied Triangles.''  In the general case, not all triangles
are satisfied, but we conjecture that the rule can be extended by
replacing ``triangles'' by more complicated spin clusters.

\noindent 3. The Rule of Satisfied Triangles implies that the
energy of a collection of non-overlapping defects is also
independent of their spatial arrangement.  Hence the system is
not a spin glass, despite the change in the degree of frustration
introduced by the disorder.

\noindent 4. For the uniform system and for moderate randomness,
we introduce a mapping of the ground
states of the kagome system onto a folded, close-packed sheet of
triangles. The folding of this sheet of ``spin triangles''
constitutes a new kind of ``spin origami,'' a term
originally coined  by Ritchey, Coleman and Chandra\cite{ref:12} for
the folding of spin planes in the kagome lattice.  The properties
of this folded sheet are related to those of ``tethered
surfaces'' which have been studied extensively by Nelson and
co-workers.\cite{ref:nelson} Folding is used to study a variety
of ground state configurations for the pure and diluted cases.

\noindent In the remainder of this letter, we justify and
elaborate on the four points listed above.

The instability of coplanar states against perturbations to the
magnitudes of the spins is easily demonstrated by the case of a
single defect spin, $S^\prime$, in an initially coplanar ground
state configuration of spins, $S$. Consider the spins in the two
hexagons which include $S^\prime$ (cf. Fig \ref{fig:1}a).
Rotations by small angles $\pm\theta$ into or out of the plane,
which alternate in sign around each hexagon, may have two
possible relative phases. If $\delta S = S^\prime - S < 0$,
then the mode which is odd under inversion through the site of
spin $S^\prime$ has a negative energy, $\delta E= JS\delta
S\theta^2$, and a node at the site of $S^\prime$.  If
$\delta S > 0$ then the symmetric mode lowers the energy by
$\delta E=-3JS\delta S\theta^2$, and $S^\prime$ is rotated by
$2\theta$.
The instability arises because there is no
quadratic restoring force for such modes in the perfect system.
For a finite density of defects, out-of-plane distortions will
compete with thermal excitations favoring coplanarity. If the
spatially averaged value of $|\delta S|/S$ is of order 1, then
thermal selection cannot supress out-of-plane canting.  If it is
small compared to 1, then the canting grows up at low
temperatures, $T/JS^2 << 1$.

Figure \ref{fig:2}a shows the results of a Monte Carlo calculation of the
low temperature nematic correlation function, $g(r)$,
which is defined in
Ref. \cite{ref:5}, for various concentrations of vacancies
between $x=0$ and $x=0.05$.  This figure shows that even a
low density of vacancies is sufficient to supress nematic order,
and suggests that nematic order should not be observable in
samples such as those discussed in Ref. \cite{ref:10}. This large
effect results from the fact, derived below, that each isolated
vacancy locks at least 10 neighboring spins into a noncoplanar
configuration.

Next we consider ground state configurations of the kagome
system, with disorder in the magnitudes of the
spins and/or dilution with vacancies, for which the nearest neighbor
Heisenberg Hamiltonian may be written as a
sum over all triangles, $\Delta$,
\begin{equation}
E=(J/2)\sum_{\Delta}\Biggl(\biggl(\sum_{m=1}^3\vec S_{\Delta,m}\biggr)^2
- \sum_{m=1}^3\vec S_{\Delta,m}^2\Biggr), \label{eqn:2}
\end{equation}
where m is summed over the three spins in a triangle.  This means that a
lower bound for the ground state energy is obtained by minimizing
the total spin of each triangle separately.  A particularly simple case
is the one in which the spins in every triangle add to zero,
i.e., where they form closed triangles, as
happens, for example, in ground states of the perfect system.

A useful device for describing kagome spin ground states is spin
origami, which is most easily visualized for the case of no
disorder.  Then in the ground state, each kagome triangle of
spins forms a closed equilateral triangle in spin space which we
call a ``spin triangle.'' Each spin forms an edge shared by
two neighboring spin triangles (See Fig. 1b.), and hence a ground
state may be represented by a folded sheet,
fashioned by joining spin triangles together along common edges.
The $\bf q$=0 state corresponds to a flat sheet (Fig.
\ref{fig:1}b). The $\sqrt{3}\times\sqrt{3}$ \ state is generated
by folding the entire sheet into a single triangle, as shown on
the left in Fig. \ref{fig:1}c.  The so-called ``weathervane
defect'' of the $\sqrt{3}\times\sqrt{3}$\ state corresponds to a
single protruding triangle, six layers thick, which can make any
angle with the rest of the spin triangles as shown on the right
in Fig. \ref{fig:1}c.

Introducing defects in the magnitudes of spins can be thought of
as shrinking or lengthening sides of spin triangles.  In general,
this will lead to buckling of the sheet. A vacancy is
generated by shrinking a bond shared by two triangles to zero
length.  The remaining sides of the triangles, which then abut,
represent antiparallel spins.  To construct the ground state for
a single vacancy by folding, a rhombus is cut out of the sheet
which is then folded so as to join the cut edges.  The resulting
sheet can be made coplanar (or even
folded into a single triangle) except for two triangles, each
four layers thick and involving the four cut edges, which
protrude out of the plane, forming two sides of a tetrahedron
(faces CEF and ADF of the tetrahedron in Fig. \ref{fig:3}). A
third face is the plane (ABC), defined by the rest of the spin
triangles. The fourth face (BDE) is empty.  In this way,
noncoplanarity can be localized to the 10 spins in the three
edges of the two protruding triangles.  (These are the 10 spins,
connected by dashed lines, in the two hexagons which share the
vacancy in Fig. \ref{fig:3}.) Of course there can also be a
larger or even infinite number of noncoplanar spins because of
the degeneracy of folding, but only the 10 spins around the
vacancy are forced to be noncoplanar. Similar arguments apply to
any isolated defect spin $S^\prime$, where, $0 < S^\prime < 2S$,
so that the surrounding spin triangles can close. For $S^\prime
> 2S$, the spins in the two
triangles involving $S^\prime$ are colinear
and the ground
state energy is equal to the lower bound in which
the sum of the spins in the two triangles involving $S^\prime$
is $S^\prime-2S$. The usual entropy
argument\cite{ref:5} implies that spins away from a vacancy will
be coplanar.  Figure \ref{fig:2}b shows the low temperature
nematic correlation function, taking the origin in a triangle
next to the vacancy. The correlation function decays in a few
lattice spacings to a value $-1/3$, which is consistent with the
spins being coplanar outside the tetrahedral arrangement of Fig.
\ref{fig:3}.

The fact that the effect of an isolated defect is localized
to  a very small area implies that, for a small
concentration of  defects, the ground state energy of the
disordered system is equal to the sum of energies of
independently satisfied triangles. In fact this statement also
applies to cases where pairs of defects are close together or
even nearest neighbors.  In the case of divacancies, the triangle
containing the two defects cannot close. Nevertheless, the ground
state obeys the rule of satisfied triangles. It can be
represented by spin origami by making a cut (a kind of
dislocation) along a line of bonds emenating from the divacancy
and then overlapping the two rows of triangles along the cut.
Furthermore, we have checked that neighboring pairs of
divacancies also satisfy the rule of satisfied triangles.  Thus
it appears that the rule applies to quite a wide range of
defect densities and configurations.

Our results for vacancies are consistent with the recent work of
Huber and coworkers\cite{ref:13} who studied the diluted kagome
antiferromagnet and obtained the striking result that the
distribution of local fields is discrete with a small number of
values: $2JS$, $JS$, and $0$.  The rule of satisfied
triangles explains this result immediately. Spins on all
triangles with no vacancies or one vacancy feel the largest local
field, and spins with two neighboring vacancies experience a
field $JS$.  The fraction of spins with this local field is
proportional to $x^2$ where $x$ is the fraction of vacancies.
Spins with zero local field arise from more complicated
configurations or from boundary effects.

It is not the case, however, that the rule of satisfied triangles
is valid for all random distributions of spins.  We have found a
number of situations in which the rule fails, all of them
involving rather strong disorder, i.e. clusters of defects. This
suggests a generalization of the rule of satisfied triangles to a
rule of satisfied clusters. The rule is simply that the ground
state energy is the sum of energies of independently satisfied
triangles and clusters.  Clusters can be identified by
calculating the local ``triangle frustration energy'' which is
the difference between the energy of a triangle in a ground state
configuration and the minimum possible energy for that individual
triangle.  Clusters are isolated regions in which the local
triangle frustration energy is nonzero.

The rule of satisfied triangles/clusters implies that the ground
state energy does not depend on the relative positions of defects
or on their degenerate degrees of freedom (such as up or down).
This is different from the usual situation in disordered spin
systems.  Two defects, which introduce frustration into a ferro-\
or antiferromagnet, induce overlapping spin distortions,
resulting in an indirect interaction between the defect degrees
of freedom and, ultimately, in spin glass behavior\cite{ref:14}. In our
system, a defect induces strong local perturbations, but the
system is so soft that this perturbation does not generate
effective pair interactions between defects. Conventional spin
glass behavior can not arise in this situation.  For a low
density of vacancies, the ground state degeneracy remains
infinite because there are still an infinite number of ways to
fold the remaining spins.  We note that similar arguments have
been made by Villain\cite{ref:14} for the case of the pyrochlore
antiferromagnet which he described as a ``cooperative
paramagnet.'' The rule of satisfied clusters breaks down when the
disorder is so strong that the clusters merge into an infinite
cluster.  In this situation, spin glass behavior may
occur.\cite{ref:14} We also note that weaker interactions which
are not included in our model, such as magnetic anisotropy,
further neighbor, interlayer and dipole couplings as well as
quantum effects\cite{ref:chubukov}, will lead to violations of
the rule of satisfied triangles and may be the source of spin
glass behavior with a very low $\rm T_f$.  It is also possible
that the apparently nonergodic behavior, which is observed
experimentally, is simply a property of the pure system at low T,
as was proposed in Ref. \cite{ref:13}.  If this is the case, then
the new type of spin origami described above will be a useful
tool for understanding the character of this low temperature
state. Conversely we note that the the spin origami mapping
seems to contain the essential ingredients for formulating a spin
model of tethered membranes of the type studied by Nelson and
co-workers\cite{ref:nelson}, thus connecting the kagome spin system
to a much broader range of physical problems.

We thank David Huber and Premi Chandra and Piers Coleman for
sending us preprints of their work.  E.F.S. is grateful for the
support of the Physics Department of the University of North
Carolina where he began working on this problem. The authors
acknowledge useful conversations and correspondence with Boris
Botvinnik, John Chalker,A. Chubukov, Malcolm Collins, Bruce
Gaulin, Chris Henley, Catherine Kallin, Steve Kivelson, David
Nelson, Art Ramirez, Jan Reimers and Byron Southern. This work
was supported by the Natural Sciences and Engineering Research
Council of Canada.

\samepage
\figure{ a) A kagome lattice of spins (dots) with nearest
neighbor interactions (solid lines). $S^\prime$ labels a defect
spin. b) The same kagome lattice of dots, superimposed on a
triangular lattice. The lengths of the sides of each triangle
represent the lengths of spins, and the arrows represent their
directions.  For simplicity, $S=S^\prime$ here. Making $S^\prime$
longer or shorter would require folding the sheet of triangles.
As shown, the figure represents the perfect $\bf q$=0 ground
state. c) (Left) A triangular sheet of $N_T$ triangles folded
into a stack.  When  flattened into a single triangle, this stack
represents the $\sqrt{3}\times\sqrt{3}$\ ground state.
(Right) The ``weather vane''
defect, a stack of $N_T-6$ triangles with six stacked triangles
(a folded hexagon) protruding.  The dihedral angle, $\phi$,
between the two stacks is arbitrary. \label{fig:1}}

\figure{ a) Nematic correlation function, as defined in Ref. 5,
for $T/J=0.00025$ and 432 lattice sites.  Solid dots are for
no vacancies ($x=0$); open dots are for 3 vacancies
($x=0.007$); the solid traingles are for 8 vacancies ($x=0.02$);
and the open triangles are for 20 vacancies ($x=0.05$).
b) Nematic correlation function $g_n({\bf r,r^\prime})$ for a
system of 431 spins plus one vacancy, with $\bf r^\prime$
located in a triangle adjacent to the vacancy and
$T/J=0.00025$.\label{fig:2}}

\figure{ Spin map of the ground state configuration around a
vacancy. A,B,C,$\pm$D,$\pm$E, and $\pm$F, denote spin orientations
in terms of the directions of labeled edges of the tetrahedron.
The tetrahedron is the spin origami representation of the same
state. The base ABC contains $N_T-8$ triangles; the faces CEF and
ADF each contain 4 triangles; and the face BDE is
empty.\label{fig:3}}


\begin{references}

\bibitem[(a)]{ref:a}  Also at Sankt Petersbourgh Nuclear Physics Institute.

\bibitem[(b)]{ref:b} Permanent address: Institute of Automation and
Electrometry, 630090, Novosibirsk, Russia.

\bibitem{ref:1} P. Fazekas and P.W. Anderson, Phil. Mag., {\bf 30},
423 (1974).

\bibitem{ref:2} X.G. Wen, F. Wilczek and A. Zee, Phys. Rev. {\bf
B39}, 11413 (1989).

\bibitem{ref:3} P. Chandra and P. Coleman, Phys. Rev. Lett. {\bf 66},
100 (1991).

\bibitem{ref:4} A.B. Harris, C. Kallin and A.J. Berlinsky, Phys. Rev.
{\bf B45}, 2899 (1992).

\bibitem{ref:5} J. Chalker, P.C. Holdsworth and E.F. Shender,
Phys. Rev. Lett., {\bf 68}, 855 (1992).

\bibitem{ref:6} J. Villain, R. Bidaux, J.P. Carton and
R. Coute, J. de Physique {\bf 41}, 1263 (1980).

\bibitem{ref:7} E.F. Shender, Sov. Phys. JETP {\bf 56}, 178 (1982).

\bibitem{ref:8} J.N. Reimers and A.J. Berlinsky (submitted to
Phys. Rev. {\bf B}).

\bibitem{ref:9} D.A. Huse and A.D. Rutenberg, Phys. Rev. {\bf B45},
7536 (1992).

\bibitem{ref:sachdev} S. Sachdev, Phys. Rev. {\bf B45}, 12377
(1992).

\bibitem{ref:10} For a brief review of this work, see A.P.
Ramirez, J.Appl. Phys. {\bf} 70, 5952 (1991) and references cited
therein.

\bibitem{ref:ram2} A.P. Ramirez, Phys. Rev. {\bf B45}, 2505
(1992).

\bibitem{ref:11} B. Mart\'inez, F. Sandiumenge, A. Rouco, A. Labarta, J.
Rodr\'iguez-Carvajal, M. Tovar, M.T. Causa, S. Gal\'i and X. Obradors,
Phys. Rev. {\bf B46}, 10786 (1992).

\bibitem{ref:henley} Destruction of order from disorder in a
system with a finite number of zero modes has been considered by
C.L. Henley, Phys. Rev. Lett. {\bf 62}, 2056 (1989) and by Ya.
Federov and E. Shender, J. Phys. Cond. Matter, {\bf 3}, 9123 (1991).

\bibitem{ref:12} I. Ritchey, P. Coleman and P. Chandra (preprint).

\bibitem{ref:nelson} See, for example, L. Radzihovsky and D.R.
Nelson, Phys. Rev. {\bf A44}, 3525 (1991) and references cited.

\bibitem{ref:13} D.L. Huber and W.Y. Ching Phys. Rev. B (in
press).

\bibitem{ref:14} J. Villain, Z. Phys. {\bf B33}, 31 (1979).

\bibitem{ref:chubukov} A. Chubukov, Phys. Rev. Lett. {\bf 69},
832 (1992).

\end{references}
\end{document}